\documentclass[twocolumn,showpacs,preprintnumbers,amsmath,amssymb,prb]{revtex4-1}

\usepackage{graphicx}
\usepackage{dcolumn}
\usepackage{bm}
\usepackage{color}

\begin{document}

\preprint{APS/123-QED}

\title{Predicting the electronic structure of weakly interacting
hybrid systems: \\ The example of nanosized pea-pod structures}

\author{Matus Milko$^{1,2}$}
\email{matus.milko@uni-graz.at}
\author{Peter Puschnig$^{1,2}$}
\author{Claudia Draxl$^{1,3}$}

\affiliation{$^{1}$Chair of Atomistic Modelling and Design of Materials, University of Leoben, Franz-Josef-Stra\ss e 18, A-8700 Leoben, Austria}
\affiliation{$^{2}$Institute of Physics, Theoretical Physics, University of Graz, Universit\"atsplatz 5, A-8020 Graz, Austria}
\affiliation{$^{3}$Institute of Physics, Humboldt-Universit\"at zu Berlin, Newtonstra\ss e 15, D-12489 Berlin, Germany}

\date{\today}

\begin{abstract}
We provide a simple scheme for predicting the electronic structure of van-der-Waals bound systems, based on the mere knowledge of the electronic structure of the subunits. We demonstrate this with the example of nano-peapods, consisting of polythiophene encapsulated in single-wall carbon nanotubes. Using density functional theory we disentangle the contributions to the level alignment. The main contribution is shown to be given by the ionization potential of the polymer inside the host, which, in turn is determined by the curvature of the tube. Only a small correction arises from charge redistributions within the domains of the constituents. Polarization effects turn out to be minor due to the cylindrical geometry of the peapods and their dielectric characteristics. Our findings open a perspective towards designing opto-electronic properties of such complex materials.
\end{abstract}


\maketitle
The creation of nano-structures with tailored properties is an emerging field, opening exciting perspectives in all kinds of research areas and technologies. Exploiting the potential of nano-technology requires a thorough understanding of the respective building blocks as well as their interplay. Only that way, specific novel features can be designed. Theory thereby plays an important role in getting deeper insight. For a quantitative analysis one needs to make use of the most sophisticated {\it ab-initio} methods, which are in most cases too involved to be applied to real systems. Hence, simple schemes to describe specific properties of a nano-structure from the knowledge of its constituents, are most valuable tools towards materials design. In this work, we propose such a scheme to predict and hence tailor the electronic structure of weakly bound nano-hybrids. We demonstrate our approach with peapods - carbon nanotubes ({\it pods}) accommodating molecules or atoms ({\it peas}) in their cavities. 

Nano-peapods have attracted scientific interest since their discovery.\cite{Smith1998} They combine the unique mechanical properties of the nanotube with the opto-electronic properties of the guest. This way, very stable hybrid materials with tunable characteristics can be formed. In most applications, the role of the guest is to alter the electronic structure of its host in order to achieve the desired features. For example, they have been proposed for opto-electronics,\cite{Yanagi2006} electron-transport devices,\cite{Shimada2002} and even for more fundamental purposes, like one-dimensional spin arrays with an outlook towards quantum computing.\cite{Simon2006} 

While nanotubes with fullerenes inside have been studied extensively over many years, organic molecules inside the cage represent a more recent topic. After the encapsulation of Zn-diphenylporphyrin,\cite{Kataura2002} tetracyano-p-quinodimethane,\cite{Takenobu2003} perylene derivates,\cite{Fujita2005} and $\beta$-carotene,\cite{Yanagi2006} the successful synthesis of nano-peapods with thiophene oligomeres was reported.\cite{Loi2010} The particular feature of the latter is their ability to emit light in the visible range of the spectrum. This provides new opportunities for opto-electronic devices as pure carbon nanotubes are infra-red emitters, while being passive in the visible range. The question immediately arising is how the electronic and hence also the optical properties of such systems can be designed and predicted prior to synthesis. 

To address this point we exemplarily investigate a series of peapods, consisting of (n,0) single-wall carbon nanotubes (SWNTs) and polythiophene representing the guest. In the following, we use the short-hand notation pT@(n,0), and $(\infty,0)$ refers to a single graphene sheet as the limiting case of a tube with infinite diameter. We point out, that these systems are purely van der Waals (vdW) bound\cite{Loi2010,Milko2011} as will be seen in more detail below.

All calculations presented here are based on density functional theory (DFT) using ultra-soft pseudo-potentials 
as implemented in the program package QUANTUM-ESPRESSO.\cite{espresso2009} A plane-wave cutoff of 30 Ry is chosen, and a total of 4 $k$ points is taken into account for the one-dimensional Brillouin zone integrations. We adopt a super-cell approach with a vacuum size of 8{\AA} to avoid interactions between adjacent nanotubes. Exchange-correlation effects are treated within the generalized gradient approximation (GGA) in the PBE parameterization.\cite{Perdew1996} For geometry relaxations, the vdW density functional (vdW-DF)\cite{Dion2004} is employed in a post-scf manner.\cite{Nabok2009} A self-consistent vdW-DF procedurefor selected systems proved the charge density not to be affected, as already found by Thonhauser for other systems.\cite{Thonhauser2007} To ensure commensurate lattice parameters along the tube axis, we double the repeat unit of the zig-zag SWNT and stretch the polymer by about 8\%. We consider two different geometric arrangements, one with the polymer centered inside the tube and another one where it is sitting at its optimal distance from the tube wall. 

\begin{figure}[h]
\begin{center}
{\includegraphics[width=7.0cm]{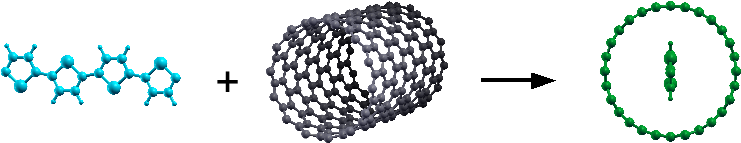}}
\end{center}
\begin{center}
{\includegraphics[width=\columnwidth]{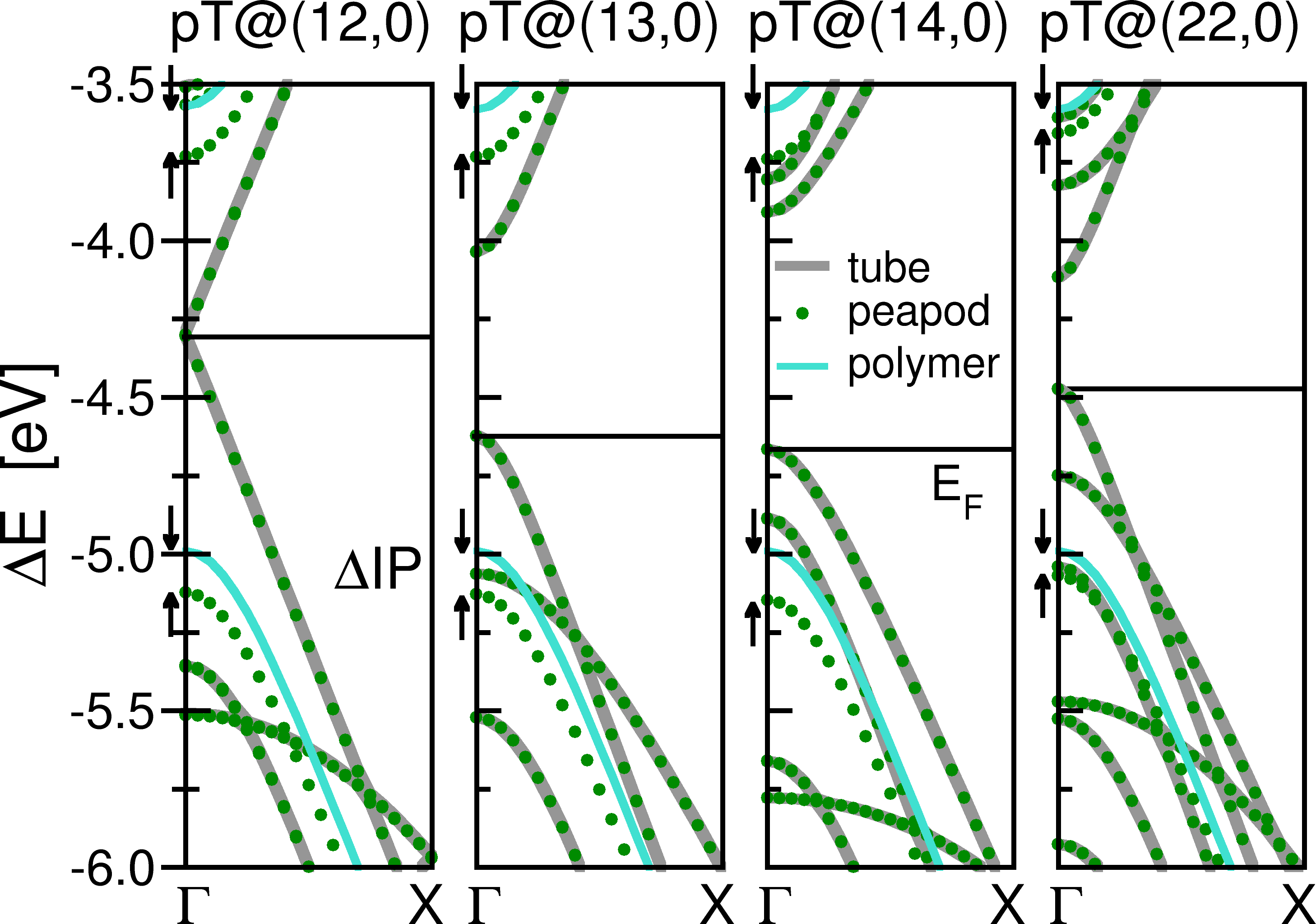}}
\end{center}
\vspace{-0.5cm}
\caption{\label{fig:bands}
Band structures of the peapods pT@(n,0) with n = (12, 13, 14, 22) shown in the vicinity of the Fermi energy (black horizontal lines). Full turquoise lines, thick gray lines, and green dots represent the electronic bands of the isolated polythiophene, isolated nanotube, and the peapod, respectively. All band energies are aligned with respect to the vacuum level. The peapod bands originating from the polymer appear to be rigidly shifted by an amount $\Delta$IP compared to the isolated polymer, as indicated by the arrows.}
\end{figure}

In Fig.~\ref{fig:bands} the band structures obtained by DFT calculations of the isolated polythiophene (turquoise lines), the isolated nanotubes (bold gray lines), and the corresponding peapods (green dots) are plotted for four zig-zag SWNTs. For any system consisting of fully non-interacting subunits, the total electronic structure is a superposition of those of its constituents, aligned with respect to the vacuum level, {\it i.e.}, the electrostatic potential at infinite distance. This so-called Schottky-Mott limit\cite{Kahn2002, Schottky1938} should generally apply to vdW bound systems, where no charge transfer between the constituents occurs. Hence, the bands in Fig.~\ref{fig:bands} are aligned in this manner. In fact, the electronic states of both subunits retain all their features when forming the band structure of the peapod. We observe, however, that those peapod bands which are originating from polymer states appear rigidly shifted downward (by an energy labeled $\Delta$IP) compared to the band structure of the isolated polythiophene. These shifts are rather small, {\it i.e.}, 0.12, 0.14, 0.15, and 0.08 eV for pT@(12,0), pT@(13,0), pT@(14,0), and pT@(22,0), respectively. In the following, we aim at explaining their origin, thereby enabling a prediction of the peapod's band structure from the knowledge of its constituents' bands alone. For this, we have to consider three different factors. The first one is specific to carbon nanotubes, the second one emerges from charge rearrangements between weakly interacting entities, and the third factor is connected with mutual polarization of the subunits. These will be discussed in the following. 

\begin{figure}[ht]
\begin{center}
\includegraphics[width=8.5cm]{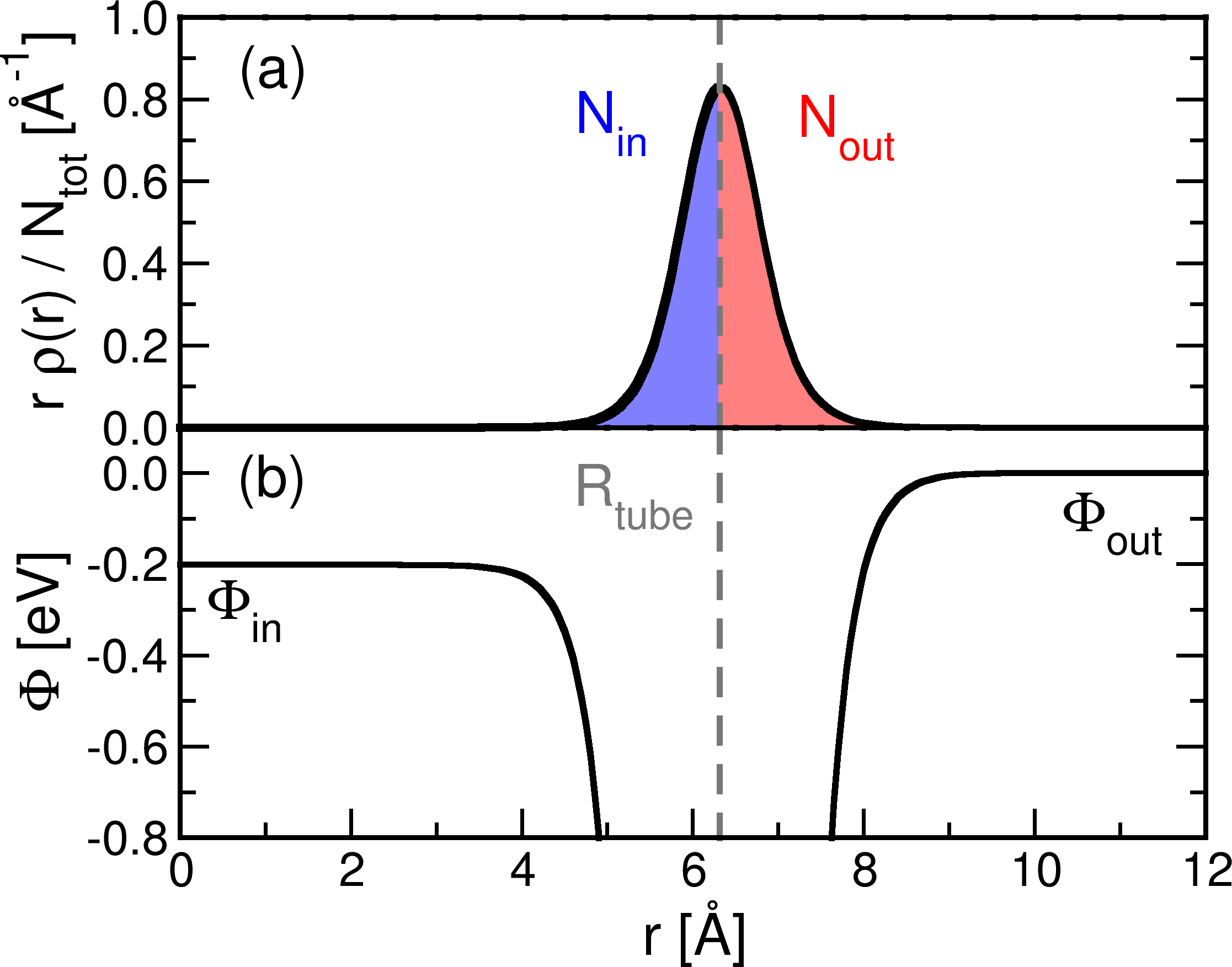} \vspace{0.2cm} \\
\includegraphics[width=8.5cm]{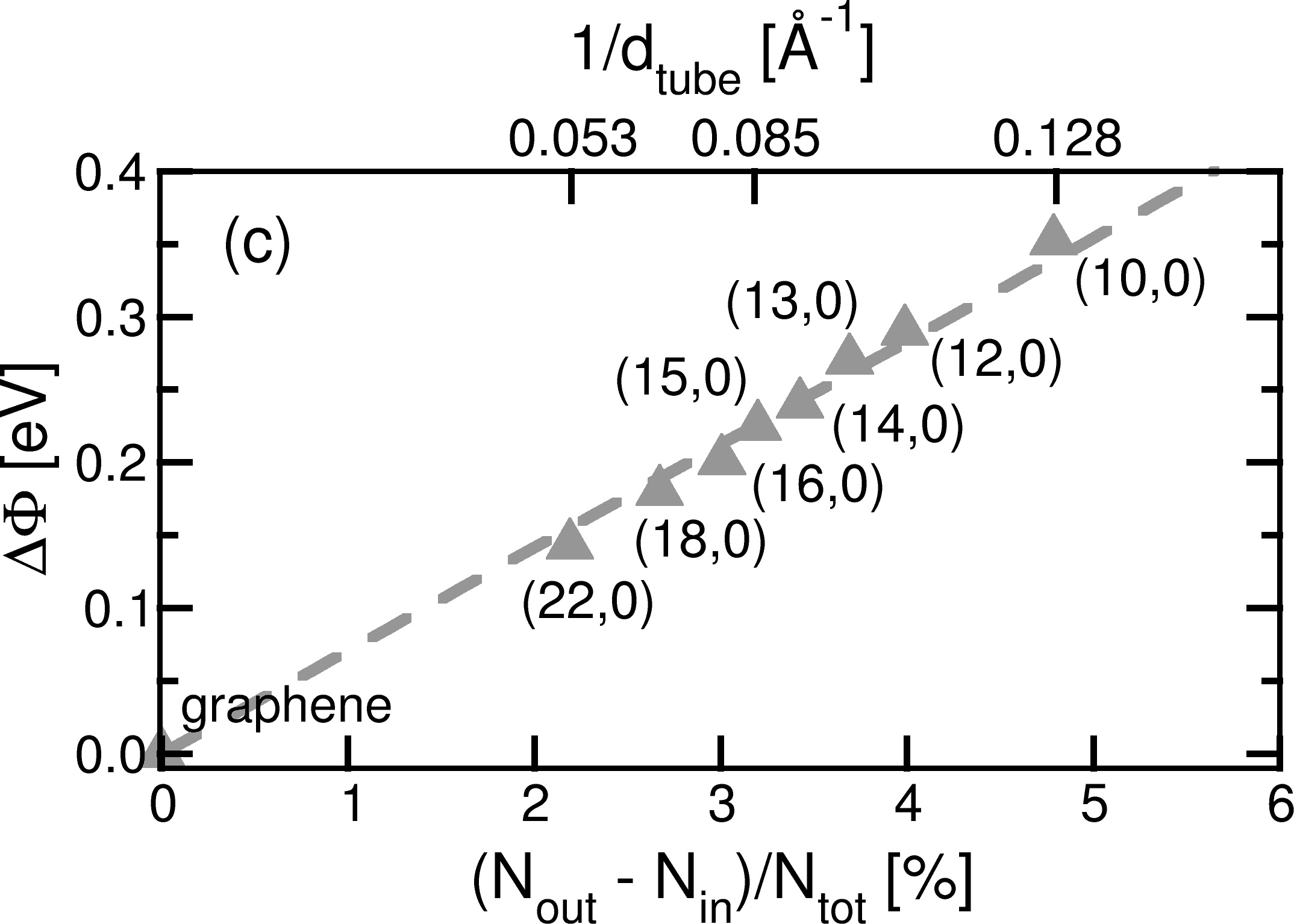}
\end{center}
\caption{(a) 
Charge density of a (16,0) SWNT as a function of the radial coordinate $r$, averaged along the tube axis and the azimuthal angle. The blue and red areas below the curve represent the charges inside and outside the tube perimeter, $N_{in}$ and $N_{out}$, respectively. (b) The corresponding electrostatic potential averaged along the tube axis. (c) Difference in the electrostatic potential inside and outside the tube perimeter as a function of the normalized charge asymmetry.}
\label{fig:asym}
\end{figure}

The first factor stems from the curvature of the carbon nanotubes. It gives rise to a charge asymmetry between the inside and outside area with respect to the tube wall. This is visualized in the top panel of Fig.~\ref{fig:asym}a. Here, the charge density, averaged along the tube axis and the azimuthal angle, is depicted for a (16,0) SWNT as a function of the distance from the tube center, $r$. The areas below this curve left and right of the tube perimeter (gray dashed line), marked in blue and red, respectively, represent the corresponding number of electrons. The observed charge asymmetry results in an offset in the electrostatic potential inside the host compared to the vacuum level. This is illustrated in Fig.~\ref{fig:asym}c where we plot the potential difference $\Delta\Phi = \Phi_{out}-\Phi_{in}$ for a series of (n,0) tubes versus $(N_{out}-N_{in})/N_{tot}$. We find $\Delta\Phi$ to depend linearly on this normalized charge asymmetry and to be proportional to the tube curvature $1/d$. As a consequence of this electrostatic effect, molecules positioned inside or outside the tube experience a different electrostatic potential, and thus the molecular levels move downwards as compared to the isolated molecule. 

\begin{figure}[ht]
\begin{center}
\includegraphics[width=\columnwidth]{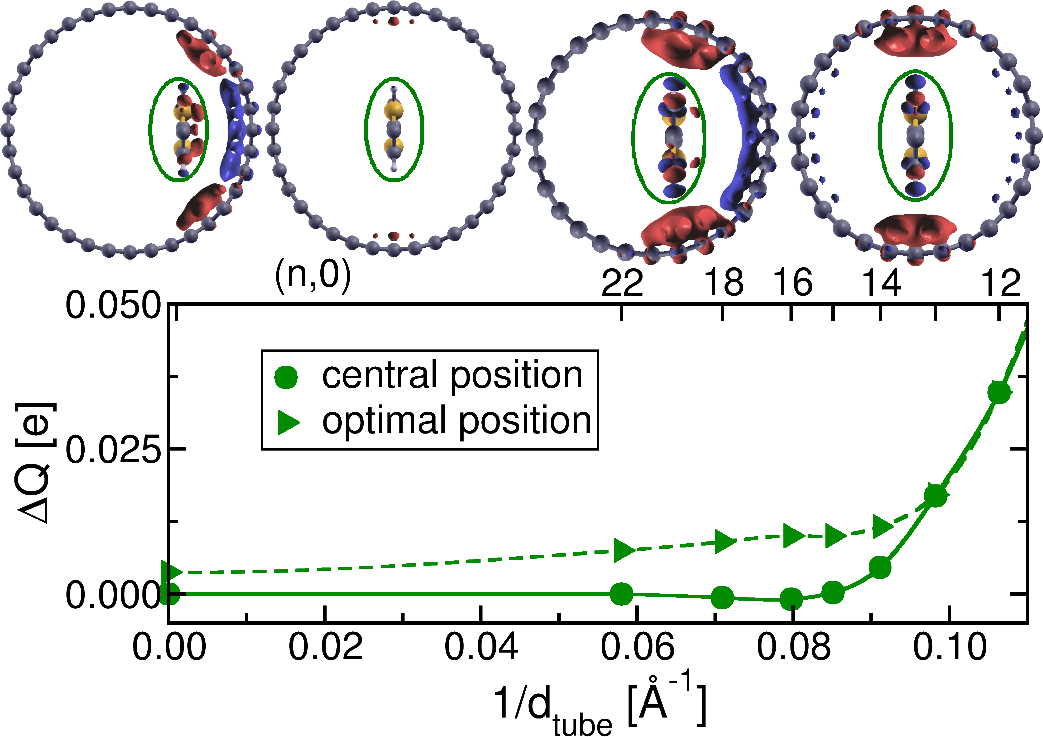}
\end{center}
\caption{\label{fig:charge}
Top: Charge density difference integrated along the tube axis with isovalues of +0.0001 (red) and -0.0001 (blue)  electrons per a.u.$^2$. Results are displayed for pT@(18,0) (two pictures at the left) and pT(14,0) 
(two pictures at the right) for central and optimal position of the polymer, respectively. The green ovals mark $\nabla\rho = 0$, separating the domains of the two subsystems. Bottom: Charge exchanged between the nanotube and the polymer in dependence of the tube diameter. The filled spheres and the triangles indicate the
cases of polythiophene in the center of the tube and its optimal distance from the tube wall, respectively.}
\end{figure}

The second factor originates from a small charge re-distribution when the guest is getting in contact with the host.
To quantify its impact, we evaluate the difference density $\Delta\rho = \rho_{peapod} - \rho_{tube} - \rho_{polymer}$, where the densities of the polymer, $\rho_{polymer}$, and that of the tube, $\rho_{tube}$, have been computed in the same geometry as that of the peapod, $\rho_{peapod}$.
This charge density difference is depicted in the upper part of Fig.~\ref{fig:charge} for pT@(18,0) (left) and pT@(14,0) (right) for the polymer in the wall and center position, respectively. 
We analyze the charge transfer between the two subunits by integrating $\Delta\rho$ separately over the regions of the polymer and nanotube, respectively, which we define through the condition $\nabla\rho_{peapod} = 0$ as indicated by the green ovals. As expected, the so obtained charge, $\Delta Q$, exhibits a systematic decrease as the tube size increases, reaching zero for the centrally situated polymer, while it approaches the tiny value of 0.004 transferred electrons in case of the optimally spaced subsystems. Hence, we can conclude that there is negligible charge passing from one subunit to the other as expected for vdW bound systems. The main charge rearrangement takes place within the individual domains leading to formation of electric dipoles 
as seen in Fig.~\ref{fig:charge}. To explore the impact of such charge dipoles quantitatively, 
we solve Poisson's equation $\nabla^{2} \Phi_{corr} = \Delta\rho$ for the differential density $\Delta\rho$. This yields another contribution,$\Phi_{corr}$, to the electrostatic potential which is analogous to the bond dipole effect discussed in the context of molecule/metal junctions.\cite{Heimel2006b}

\begin{figure}[h]
\begin{center}
{\includegraphics[width=8.6cm]{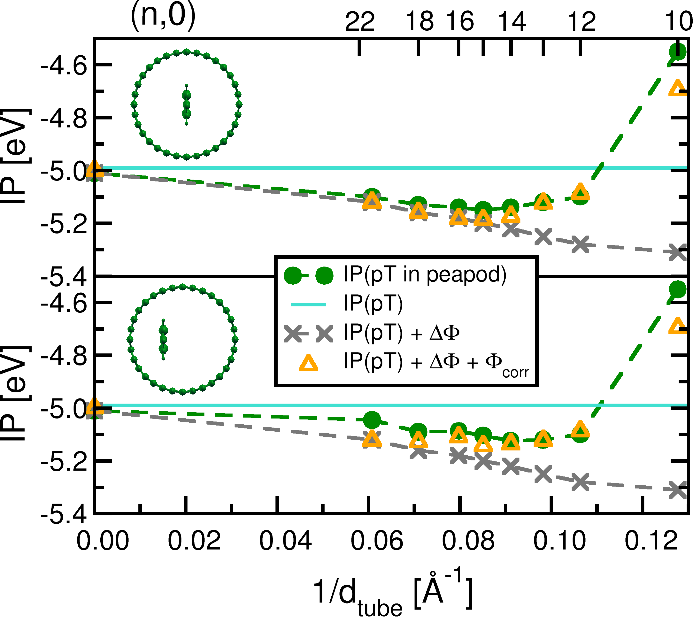}}
\end{center}
\caption{\label{fig:IP}
Ionization potential as a function of the inverse tube diameter for the polymer in the center of the tube (top) and its optimized position (bottom). The green filled symbols indicate the IP obtained by DFT calculations for the peapod.  The turquoise horizontal lines depict the IP of the isolated polymer. Gray crosses show the IP of the polymer inside the cavity at its respective position. The orange open symbols mark the corresponding values when accounting for the charge dipoles arising between the polymer and the tube.}
\end{figure}

The results obtained so far are summarized in Fig.~\ref{fig:IP}. Here, the ionization potential (IP) of the polymer inside the SWNT as a function of the tube diameter is displayed for the two different positions of the pea inside the pod. The turquoise lines correspond to the IP of the isolated polymer, and the green filled circles represent the DFT results for the peapod. Their difference is identical to the quantity $\Delta$IP indicated by the arrows in Fig.~\ref{fig:bands}. It provides an estimate for the accuracy we can expect when considering the Schottky-Mott limit as a zeroth approximation to the level alignment. The two factors discussed above represent corrections to this picture. First, the fact that the polymer experiences a different IP inside the pod compared to outside, gives rise to the correction $\Delta\Phi$. The resulting data are marked by gray crosses. Including this correction, determined by the tube curvature, leads to very good agreement with the results for the combined system for realistic tube diameters as accomplished experimentally.\cite{Loi2010} The deviation is within the range of 0.1eV. This is an exciting result as only the knowledge of the pristine systems is required to obtain this approximate IP. 

Evaluating these approximate IPs for the central and the off-centered position further, the deviations from the peapod data are somewhat larger for the latter. This can be understood in terms of guest-host interaction, which is also responsible for the pronounced discrepancies found for very small tubes. Here, the corrections arising from the charge dipoles, $\Phi_{corr}$, come into play. Taking them into account (orange triangles), leads to excellent agreement with the DFT results for both types of positions and the entire curvature range. Obviously, for very narrow tubes, the charge rearrangement upon encapsulation is dominant over the curvature effect. One should point out, however, that the formation of such peapods is energetically unfavorable.\cite{Loi2010, Milko2011}

At this point, we should assess the reliability of DFT in describing such scenarios as discussed here and exploiting its predictive power. To start with, we point out again that we are dealing with vdW-bound systems. It is clear that conventional (semi)local functionals can fail badly in this context. However, vdW-DF\cite{Dion2004} based on the PBE density has proven successful for the systems under investigation\cite{Nabok2008, Loi2010} without leading to spurious charge transfer. Concerning the DFT band-gap problem, we point out that for our analysis we do not make use of band gaps. We only evaluate the valence band structures to demonstrate the principles behind the alignment of the respective HOMO levels. The suggested procedure can equally be based on calculations performed with hybrid functionals or the {\it GW} approach, or even solely on experimental data.       

Evaluating the third factor, however, which is connected with the polarization response of the SWNT upon adding a hole (electron) in the polymer's HOMO (LUMO), (semi)local DFT would certainly fail. We emphasize, however, that this effect is small and, can equally be understood in terms of electrostatic considerations. Such scenario has been discussed for the example of benzene physisorbed on graphite,\cite{Neaton2006} showing that the $GW$ band gap of benzene is substantially reduced upon adsorption, which could be well interpreted by a simple model based on an electrostatic image potential. To quantitatively estimate these effects for our system, we adopt two different electrostatic models representing two limiting cases.

\begin{figure}[h]
\begin{center}
\includegraphics[width=\columnwidth]{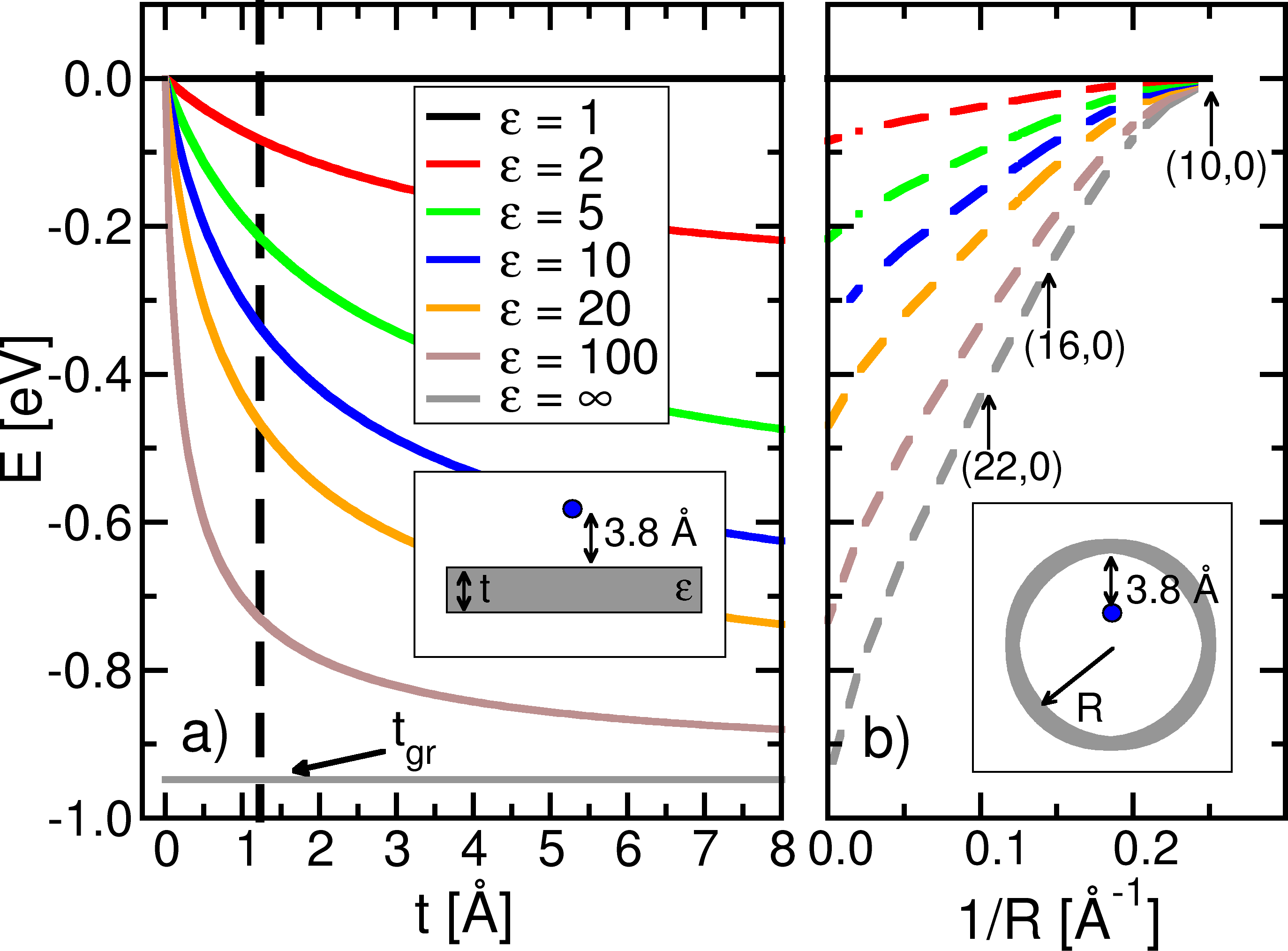}
\end{center}
\caption{\label{fig:el_stat}
 (a) Image charge model of a dielectric slab in the presence of an elementary charge: Electronic energy shift as a function of slab thickness $t$ for various dielectric constants $\varepsilon$. The vertical dashed line at $t_{gr}$ indicates the effective thickness of a graphene sheet. (b) Image charge model of an infinite metallic cylinder hosting an elementary charge placed at a distance of 3.8 $\AA$ from the cylinder wall: Electronic energy shift as a function of the inverse cylinder radius $1/R$. In both graphs the same color code for the dielectric constants is used.}
\end{figure}
 
The first model\cite{Barrera1978} considers a point charge at the equilibrium polymer-SWNT distance of around 3.8 {\AA} in front of  a dielectric slab of thickness $t$. The results are shown in Fig.~\ref{fig:el_stat}a, where the electron energy $E$ of the point charge with the induced image charge is plotted as a function of $t$ for various dielectric constants $\varepsilon$ of the slab material. Approximating the thickness of graphene by its electron density's full-width-at-half-maximum value, represented as vertical dashed line, we observe a strong dependence of the image-potential effect on the dielectric properties of the slab. A perfect metal ($\varepsilon=\infty$) would result in level shifts of almost 1 eV -- independent of the slab thickness. This corresponds well to $GW$ results for the frontier molecular orbitals of benzene adsorbed on graphite.\cite{Neaton2006} (Note that in these calculations, the adsorption distance was much smaller, 3.25 $\AA$.) However, the SWNTs studied in this work are less polarizable (e.g.~$~5$ for (14,0) tube), and, therefore, polarization shifts of about 0.1-0.3 eV only are expected (see Fig.~\ref{fig:el_stat}a). 

The above estimates are based on a planar dielectric slab. In order to take the curvature of the SWNT into account, we employ a second model where a point charge is located inside an infinitely long metallic cylinder,\cite{Hernandes2005} 
and evaluate the energy $E$ as a function of inverse cylinder radius. As follows from Fig.~\ref{fig:el_stat}b, metallic cylinders of radii corresponding to (22,0), (16,0), and (10,0) SWNTs would give rise to polarization shifts of -0.4 eV, -0.2 eV, and 0 eV, respectively (gray line). Note that for (10,0) the charge resides in the center of the tube where no polarization shift is to be expected to due symmetry reasons. As, however, the limit of metallic cylinders present an upper bound to the polarization shift, a considerable reduction is to be expected, by considering the finite dielectric constant of the SWNT. Unfortunately, no analytical solution is available for a point charge off the axis inside cylindrical dielectric of finite thickness. In order to estimate the reduction to the finite $\varepsilon$ starting from the exact results for a perfect metallic cylinder, we therefore proceed as follows. We adopt the polarization shifts from the previous model (point charge in front of a planar dielectric slab) and include these values also in Fig.~\ref{fig:el_stat}b for zero curvature ($1/R = 0$), \emph{i.e.}, the planar case. We further assume that the curvature dependence for finite  dielectric constant exhibits the same behavior as that for the perfect metal. With this extrapolation, we can now estimate the image charge effect for arbitrary nanotube diameter and given dielectric constant as presented by the dashed curves in Fig.~\ref{fig:el_stat}b. This reasoning allows us to estimate the overall polarization effects to be in the order of 0.1 eV or less for SWNT diameters and $\varepsilon$-values relevant for our examination here.

To summarize, we have discussed the level alignment of weakly bound nano-peapods. We find that due to the vdW bonding, there is negligible charge transfer between the subunits, preserving the nature of the single bands within the electronic structure of the combined system. The alignment of the HOMO levels is governed by the ionization potentials of the separated entities with a small correction arising from charge rearrangements induced by their proximity. It is important to consider the IP of the pea inside the tube, accounting for the different electrostatic environments inside and outside its perimeter. These IPs are uniquely determined by the tube curvature. We provide them for a series of SWNTs in this work and encourage their measurement, {\it e.g.}, by core level or photoemission spectroscopy. Polarization could be an important factor und needs, in general, to be taken into account. However, due to the cylindrical symmetry of nano-peapods and their dielectric properties, the corresponding electronic energy shifts are considerably reduced.

Our findings allow for predicting the band structure of nano-hybrids from the mere knowledge of its constituents. From our examples, which are representative for molecular systems used in opto-electronics, we estimate the error bar to be of the order of 0.1 eV. Such predictions could be extremely useful for designing electronic properties of nano-structures towards opto-electronic devices or nano-transistors where tuning the electronic levels is crucially important.

\bigskip
Work carried out within the NanoSci-ERA project NaPhoD (Nano-hybrids for Photonic Devices), supported by the Austrian Science Fund, projects I107 and S9714. 


%
\end{document}